\documentclass[
 reprint,
 amsmath,amssymb,
 aps,
 pre,
 floatfix
]{revtex4-2}

\usepackage{graphicx} % Include figure files
\usepackage{dcolumn}  % Align table columns on decimal point
\usepackage{bm}       % Bold math
\usepackage{babel}

\usepackage{xcolor} 

\begin{document}

\title{\textbf{A \NEW{model for} the dimension of the inertial manifold of the Kuramoto-Sivashinsky equation}}

\author{Melisa Y. Vinograd}
 \affiliation{Universidad de San Andrés, Buenos Aires, Argentina}
  \affiliation{Departamento de Física, Universidad de Buenos Aires, CABA, Argentina}

\author{Patricio Clark Di Leoni}
 \email{Contact author: pclarkdileoni@udesa.edu.ar}
\affiliation{
Universidad de San Andrés, Buenos Aires, Argentina
}%
\affiliation{
CONICET, Argentina
}%

\newcommand{\CITE}[1]{\textcolor[HTML]{E76F51}{\textbf{[Citation needed]}}}
\newcommand{\MELISA}[1]{\textcolor[HTML]{2A9D8F}{\textbf{[Melisa: #1]}}} 
\newcommand{\PATRICIO}[1]{\textcolor[HTML]{D1495B}{\textbf{[Patricio: #1]}}} 
% Changes from the first round of review (no longer highlighted;
% restore \textcolor{blue}{#1} to show them again)
\newcommand{\REFtwo}[1]{#1}
\newcommand{\REFone}[1]{#1}
% Changes in response to the second report of the second Referee
%\newcommand{\NEW}[1]{\textcolor{red}{#1}}
\newcommand{\NEW}[1]{#1}

% latex
\newcommand{\Rey}{\mathrm{Re}}
\newcommand{\fref}[1]{\figurename~\ref{#1}}

\date{\today}

\begin{abstract}
We \NEW{propose a model, based on a} spectral energy balance argument, \NEW{that provides an estimate of} the dimension of the inertial manifold of the one-dimensional Kuramoto–Sivashinsky equation (KSE). By identifying the wavenumber at which energy injection and dissipation balance out, we obtain a cutoff scale $k_d=\sqrt{5/(3\nu)}$ that we use to calculate the number of dynamically active Fourier modes. This leads to an explicit expression \NEW{$d_M(L,\nu)$ for the dimension of an effective manifold spanned by these modes, which we take as our estimate of the inertial manifold dimension and} which scales extensively with domain size. Direct numerical simulations \NEW{are consistent with the estimated} spectral cutoff and its scaling with viscosity. Using convolutional autoencoders as well as Lyapunov-based estimates, we show that the minimal latent dimension required for accurate reconstruction coincides with $d_M$ across a range of $(L,\nu)$. 
\end{abstract}

\maketitle

% ----------
% Main text 
% ----------

\section{Introduction}

Extended chaotic dissipative systems tend to exhibit finite-dimensional dynamics, despite formally living in infinite-dimensional spaces.
Systems described by the Kuramoto-Sivashinsky equation, the complex Ginzburg-Landau equations, or the Navier-Stokes are all examples of this behavior. For the first two of these examples it can actually be proven
that long time solutions live inside an inertial manifold of finite dimension \cite{foias_inertial_1988,robinson_Finitedimensional_1995}, but quantifying the dimension of the inertial manifold (or of related finite manifolds) and generating accessible representations still remains a challenge which garners considerable attention today \cite{linot_data-driven_2022,Vinograd_Clark}. In this work we focus on the Kuramoto-Sivashinsky equation (KSE). In this case, loose upper bounds estimate that the dimension scales as $L^{2.65}$ \cite{robinson_Finitedimensional_1995}, with $L$ being the domain size, \REFone{and sharper rigorous bounds were obtained subsequently~\cite{jolly_evaluating_2000}}. Further progress has been made through the analysis of covariant Lyapunov vectors, their exponents and the structure of unstable periodic orbits in the system \cite{yang_hyperbolicity_2009,takeuchi_hyperbolic_2011,ding_estimating_2016}, \REFone{with comprehensive Lyapunov spectra later computed over a wide range of $L$~\cite{edson_lyapunov_2019}}, where strong evidence of a robust way to numerically estimate the dimensionality of the inertial manifold was presented, along with the corresponding results. These estimates showed that the dimension scaled linearly with domain size, but no explanation was put forth as to why this was the case. Interestingly, the authors also showed how the Kaplan-Yorke dimension \cite{kaplan_chaotic_1979} is consistently lower than their estimates by a factor of almost two.
Later, it was shown that estimates of the dimension of the inertial manifold can also be obtained through neural-network-based autoencoder algorithms\REFone{~\cite{linot_deep_2020, sondak_learning_2021}}. These were not only able to recover the previously obtained results, but also generate representations that can be successfully described by ordinary differential equations \cite{linot_data-driven_2022}. \REFone{Data-driven approaches, including reservoir computing and recurrent neural networks, have also been used to forecast the KSE and to estimate its Lyapunov spectrum from learned models~\cite{pathak_using_2017, pathak_model-free_2018, vlachas_data-driven_2018, vlachas_backpropagation_2020}.}

In this manuscript we \NEW{propose a model that provides an estimate of} the dimension of the inertial manifold of the KSE in terms of the system parameters \NEW{and offers an explanation for} previously observed numerical findings. \NEW{The argument, based on a spectral energy balance, is not a rigorous derivation and does not yield a bound on the dimension of the inertial manifold in the mathematical sense. Rather, it estimates the dimension of an effective manifold spanned by the dynamically active Fourier modes, which can be compared with the numerical estimates of the inertial manifold dimension available in the literature.} Using an autoencoder-based representation learning approach, as well as Lyapunov-based analysis, we \NEW{show} that the minimal latent dimension necessary for accurate reconstruction \NEW{agrees with} the number of dynamically active degrees of freedom \NEW{estimated by our model}.

\section{Derivation of the model}

\begin{figure}
    \centering
    \includegraphics[width=0.95\linewidth]{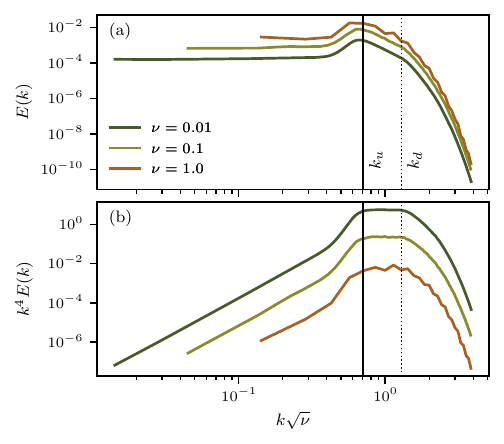}
    \caption{
    (a) Energy spectra $E(k)$ and (b) weighted spectra $k^{4} \; E(k)$ for KSE with $L=44$ and viscosities $\nu = 0.01,\,0.1,\,1.0$. 
    The spectra are plotted as a function of the rescaled wavenumber $k\sqrt{\nu}$. The vertical solid and dotted lines indicate the characteristic scales $k_{\mathrm{u}}$ and $k_{\mathrm{d}}$, respectively.    }
    \label{fig:energy_spectrum}
\end{figure}

\begin{figure}
    \centering
    \includegraphics[width=\linewidth]{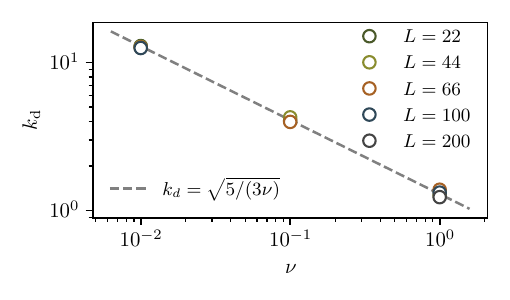}
    \caption{Dissipation cutoff wavenumber $k_d$ as a function of viscosity $\nu$. 
    The dashed line shows the \NEW{energy-balance estimate} $k_d=\sqrt{5/(3\nu)}$. Symbols indicate the numerically identified end of the approximately flat region in the dissipation spectrum $k^4E(k)$ for different domain sizes $L$. The numerical estimates collapse onto this scaling across parameters. }
    \label{fig:kd_prediction}
\end{figure}

The KSE reads

\begin{equation}
    \frac{\partial u}{\partial t} + u \frac{\partial u}{\partial x} = \NEW{-}\frac{\partial^2 u}{\partial x^2} - \nu\frac{\partial^4 u}{\partial x^4} ,
\end{equation}
where $\nu$ is the viscosity and $u$ has periodic boundary conditions in a domain of size $L$. Let us recall some known facts about the KSE dynamics\REFone{~\cite{holmes_turbulence_1996, wittenberg_scale_1999}.} \REFone{Energy is injected into the system via the $\partial^2_{xx}u$ term, while energy is dissipated via the $\nu \partial^4_{xxxx}u$ term. The nonlinear term is conservative and only redistributes energy between modes. Its dynamics are characterized by the creation and destruction of structures and the energy reaches a statistically stationary state.} In \fref{fig:energy_spectrum} we show the energy $E(k)$ (top panel) and dissipation $k^4 E(k)$ (bottom panel) spectra, obtained from numerical simulations performed with $L=44$ and various values of $\nu$, more details on the numerical simulations are given below. At low wavenumbers both production and dissipation of energy are negligible, so modes are thermalized and the energy is equipartitioned, thus showing a flat spectrum. As energy injection increases, the spectrum rises, reaching a peak near the most linearly unstable mode, corresponding to $k_u = \sqrt{1/(2\nu)}$. This mode is indicated with the vertical solid line. Beyond this peak, the system shows a short ``inertial'' range, until dissipation becomes too strong and decay becomes exponential. While $k_u$ has received considerable attention in the literature, the wavenumber at which dissipation begins to dominate has been, for the most part, neglected. We now propose a prediction for this wavenumber. This marks our first major result and will form a central part of the argument for the \NEW{estimate} of the dimension of the inertial manifold.  We propose that this transition occurs at the scale $k_d$ defined by the balance between the injection of energy and dissipation such that

\begin{equation}
    \int^{k_d}_0 (k^2 - \nu k^4) \; dk = 0,
\end{equation}
which yields $k_d =\sqrt{5/(3\nu)}$.  This wavenumber, marked by the vertical dotted line in both panels of \fref{fig:energy_spectrum}, coincides with the end of the inertial range and the onset of exponential decay. In the dissipation spectrum $k^4E(k)$, this wavenumber delimits the inertial range: the spectrum remains approximately flat for $k_u \lesssim k \lesssim k_d$, and decays rapidly for $k>k_d$. We checked this in a systematic way by performing simulations with different values of $L$ and $\nu$ and identifying the wavenumber at which the dissipation spectrum $k^4E(k)$ departs from its flat plateau. We define the plateau level $P$ as the mean of $k^4E(k)$ over the active band $k_u \lesssim k \lesssim k_d$, and take the numerical cutoff as the wavenumber where $k^4E(k) = (1-p)\,P$, with $p = 0.05$; the estimate shifts by less than one Fourier mode for $p$ between $0.05$ and $0.1$. The resulting numerical estimates, shown in \fref{fig:kd_prediction}, collapse onto the predicted scaling \NEW{of the model} for $k_d$ across the range of parameters considered. Details of the simulations are provided below.

Before moving on, we want to point out that although we carry out our calculations in the continuous approximation, we show in Appendix~\ref{appA:discrete_modes} that evaluating the exact discrete sum over the Fourier modes shifts the \NEW{estimate} of $k_d$ by at most half a Fourier mode ($\approx \pi/L$). This results in a relative error of $10.7\%$ for our coarsest setting ($L=22, \nu=1.0$) that rapidly decays for all other configurations. We thus retain the continuous approach.

Given this picture the fact that estimates of $d$ scale linearly with $L$ follows naturally. If $\nu$ is fixed, and as both $k_u$ and $k_d$ are independent of $L$, increasing the spatial domain only increases the number of thermalized modes, without altering the characteristic dynamical scales. Therefore adding new and independent degrees of freedom to the system in a linear fashion. So the dimension grows extensively $d \propto L$, as seen in previous results \cite{yang_hyperbolicity_2009,linot_data-driven_2022} \REFone{and as expected for spatiotemporal chaos in extended systems~\cite{cross_pattern_1993}.}

We now move forward with a quantitative \NEW{estimate of} the dimension of the inertial manifold. \NEW{The model rests on a simple assumption}: given that the KSE reaches a stationary energy and that the nonlinear term is conservative, the system needs enough modes to \textit{linearly} balance the injection and dissipation of energy. \NEW{We therefore assume that the dynamics requires} all modes with $|k|\le k_d$. This argument mirrors the one behind the Kaplan-Yorke conjecture, where the dimension of the system is determined by the number of modes required to balance the expansion and contraction of the phase space, as well as, resembles that used in turbulence where the number of degrees of freedom is given by the number of modes below the Kolmogorov scale \cite{landau_fluid_2013}. \NEW{Within this model, the dimension $d_{M}$ of the effective manifold spanned by these active modes, which we take as our estimate of the dimension of the inertial manifold, is} given by the number of Fourier modes in this spectral band. Counting modes in a periodic domain of size $L$ gives

\begin{equation}
    d_{M} = 2 \frac{L}{2\pi} k_d = 2 \frac{L}{2\pi} \sqrt{\frac{5}{3\nu}},
\end{equation}
where $(L/2\pi) \; k_d$ is the number of Fourier modes up to wavenumber $k_d$ and the prefactor $2$ comes from the fact that each Fourier mode contributes two real degrees of freedom due to its complex nature, while remembering that the reality condition does away with the negative wavenumbers. For $\nu=1.0$, this expression \NEW{gives} $d_M=9, 18,$ and $27$ for $L=22, 44,$ and $66$, respectively, in close agreement with previously reported values $d_M=8, 18,$ and $28$~\cite{yang_hyperbolicity_2009, linot_data-driven_2022}, differing by at most one mode. \NEW{We stress that this is a heuristic argument: it is not derived from the rigorous theory of inertial manifolds~\cite{foias_inertial_1988,robinson_Finitedimensional_1995} and does not provide a bound on their dimension. Its value lies in giving a simple, parameter-explicit estimate that can be tested against independent numerical estimates, which we do in the next section.}

These previous estimates of the inertial manifold dimension of the KSE relied on equation-based techniques such as covariant Lyapunov vectors and Floquet analyses of unstable periodic orbits \cite{yang_hyperbolicity_2009, takeuchi_hyperbolic_2011,
randons_geometry_2012, ding_estimating_2016}, primarily with $\nu=1.0$, with \citet{yang_hyperbolicity_2009} extending the analysis up to $L=196$. These estimates are obtained by computing the full Lyapunov spectrum and locating the index at which it steeply drops from near-zero exponents into strongly negative ones; the modes below this drop are identified as the dynamically active, or physical, modes, and their count is the reported dimension. More recently, autoencoder architectures have been used to infer the intrinsic manifold dimension directly from data by monitoring the reconstruction error as a function of bottleneck size \cite{linot_deep_2020, linot_data-driven_2022, zeng_autoencoders_2024}. In these studies, also with $\nu=1.0$, the dimension identified agrees quantitatively with the values obtained from Lyapunov and Floquet analyses. Beyond the KSE, autoencoders have been used to construct reduced representations of complex fluid flows and to investigate attractor dimensionality in related systems \cite{Vinograd_Clark, cleary_estimating_2025, vinograd_dimensional_2026, vlachas_multiscale_2022}. In the following section we provide further numerical validation of our model using the autoencoder and Lyapunov-based techniques mentioned above, and third hybrid autoencoder-Fourier method we propose. We recover the past estimates quoted previously and provide new ones for different values of $L$ and $\nu$, as well as provide insights into the nature of the inertial manifold.

\section{Numerical validation}

Direct numerical simulations of the KSE were performed for domain sizes $L=22, 44, 66, 100,$ and $200$, and viscosities $\nu=1.0, 0.1,$ and $0.01$, at spatial resolutions $N=256, 512$ or $1024$ depending on $L$, chosen to ensure adequate spectral resolution of the energy and dissipation ranges. For each pair $(L,\nu)$, solutions were computed using a pseudo-spectral solver with second-order Runge–Kutta time stepping from 10 distinct initial conditions.

\subsection{Autoencoder-based estimates}

The aim of an autoencoder is to approximate the identity operator with the caveat that at some point of the network there should be a bottleneck layer of size $d$ with $d$ smaller than the size of the input data. The autoencoders were implemented using convolutional neural networks and no regularizations were used in the latent layer. Implementation details and network architectures are provided in the accompanying code repository \cite{repo_paper}. For the training, validation and testing we created independent datasets of $\mathcal{O}(10^{5})$, $\mathcal{O}(10^{4})$, $\mathcal{O}(10^{4})$, respectively. We tested different values of $d$ for every pair of $(L, \nu)$. For each $d$, we trained separate networks by minimizing a mean–squared reconstruction loss with the Adam optimizer (initial learning rate $7.5\times10^{-4}$, batch size 16), reducing the learning rate by a factor of 0.75 whenever the validation loss failed to improve over 2 consecutive epochs (down to a floor of $10^{-6}$) and retaining the checkpoint with the lowest validation loss. Training was run until the validation loss plateaued, this was typically within $\sim50$ epochs, with the loss changing by less than $3\%$ over the final 10 epochs confirming that the networks had converged. The numerical estimate $d_A$ is defined as the smallest bottleneck dimension for which the validation error reaches a plateau, indicating that the latent space is sufficient to parameterize the dynamics, a criterion similar to those presented in the references above. 

\begin{figure}
    \centering
    \includegraphics[width=0.95\linewidth]{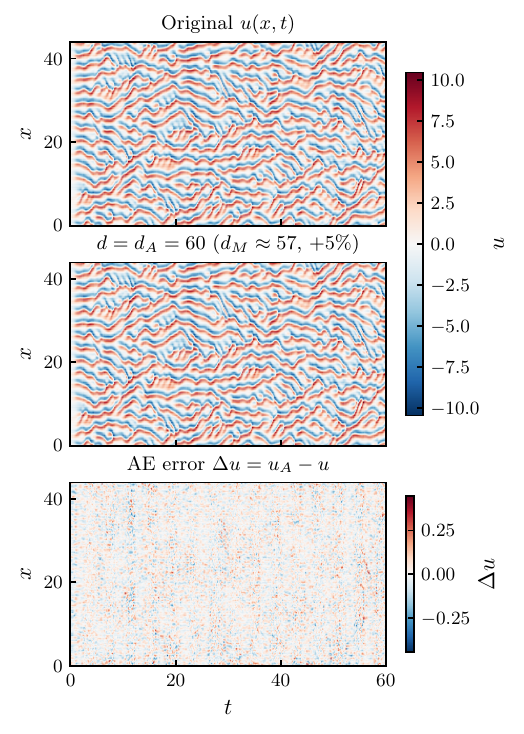}
    \caption{\REFone{Space-time reconstruction for $L=44$, $\nu=0.1$. \textbf{Top:}
    original field $u(x,t)$. \textbf{Middle:} autoencoder reconstruction at
    $d=d_A=60$ in close agreement with $d_M\approx57$. \textbf{Bottom:}
    pointwise error $\Delta u = u_A - u$.} }
    \label{fig:spacetime_L44_nu0.1}
\end{figure}

\begin{figure*}
    \centering
    \includegraphics[width=0.95\linewidth]{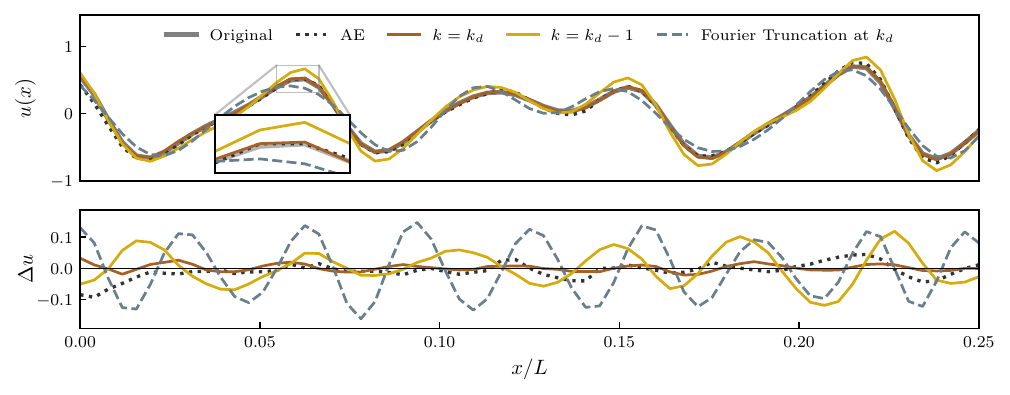}
    \caption{
    Reconstruction of a representative field for $L=44$ and $\nu=0.1$.
    \textbf{Top:} original signal $u(x)$ compared with three reduced representations: Fourier encoder with $k_c\leq k_d - 1 = 28$ and $k_c\leq k_d = 29$ and the complete autoencoder with bottleneck size $d=60$.
    \textbf{Bottom:} corresponding reconstruction errors $\Delta u(x)$.
    All curves correspond to the same snapshot.
}
    \label{fig:reconstruction_L44_nu0.1}
\end{figure*}

\fref{fig:spacetime_L44_nu0.1} shows the reconstruction in detail for the representative case $L=44$, $\nu=0.1$: the original field $u(x,t)$, the autoencoder reconstruction at $d=d_A$ and the pointwise error $\Delta u(x,t)$. The reconstruction reproduces the space-time dynamics, including the merging and splitting of structures and the drift of traveling pulses. The error stays small and spatially uniform, roughly two orders of magnitude below the field amplitude, with no concentration on particular events or coherent structures.
A single-snapshot view of the same case is shown in \fref{fig:reconstruction_L44_nu0.1}. The full autoencoder (dotted) at $d=60$ overlaps the original signal almost everywhere, with a pointwise error well below the field amplitude. The remaining curves correspond to Fourier-based encoders and are discussed in Sec.~\ref{subsec:fourier_encoder}; here we use the figure only to show the quality of the autoencoder reconstruction.
Analogous figures obtained for $\nu=1.0$ and $\nu=0.01$ are given in Appendix~\ref{appB:further_examples}.

\begin{figure}
    \centering
    \includegraphics[width=1\linewidth]{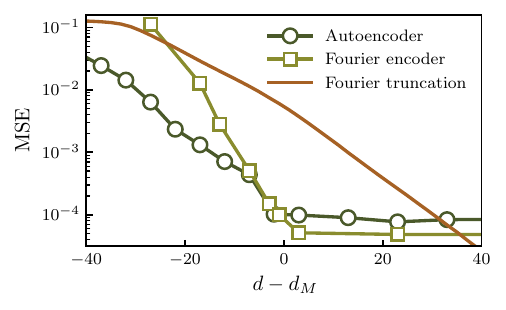}
    \caption{Mean–squared error (MSE) on the test data as a function the dimension for $L=44$ and $\nu=0.1$, where $d_M$ denotes the \NEW{estimate from our model} ($d_M=57$ for this case). Results are shown for the nonlinear autoencoder, the Fourier encoder with a learned decoder, and direct Fourier truncation.}
    \label{fig:mse}
\end{figure}

\begin{figure}
    \centering
    \includegraphics[width=0.95\linewidth]{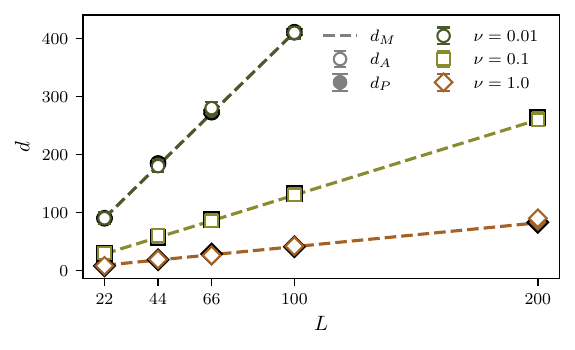}
    \caption{
        \REFone{Dimensional estimates as a function of domain size $L$ for three viscosities $\nu=0.01$, $0.1$, and $1.0$. Open markers: autoencoder
        estimates $d_{A}$. Filled markers: physical-mode counts ($d_{P}$) read
        from the drop of the Lyapunov spectrum. The dashed line is the \NEW{estimate} $d_M=(L/\pi)\sqrt{5/3\nu}$.}
    }
    \label{fig:d_vs_L_for_nu}
\end{figure}

The plateau that establishes $d_A$ is clearly identifiable from the MSE-versus-$d$ curves, shown for the representative case $L=44$, $\nu=0.1$ in \fref{fig:mse}. To automatize the procedure of finding the plateau, we define $d_A$ as the smallest sampled $d$ for which $\log_{10}\mathrm{MSE}(d)$ falls within $5\%$ of the plateau floor. This transition is universal across parameters: Appendix~\ref{appB:further_examples} shows the MSE, normalized by its plateau value, as a function of $d-d_M$ for all $(L,\nu)$ studied, collapsing onto a sharp drop near $d=d_M$. The procedure was repeated with three independent initializations of the networks and variations in filter sizes and data splits to ensure robustness\REFone{, and $d_A$ is reported as the mean over these configurations with the spread as the error bar.}
The resulting values of $d_A$ are shown in \fref{fig:d_vs_L_for_nu} alongside the predicted $d_M$, \NEW{showing close} agreement across all parameters considered. The uncertainty in $d_A$ reflects both network-to-network variability and the discrete sampling of bottleneck dimensions. In particular, the extensive scaling $d \propto L$ \NEW{given by our model} is clearly observed across domain sizes. For $\nu=1.0$, the slope \NEW{given by our estimate}, $m \approx 0.411$, agrees closely with the slope $m \simeq 0.421$ extracted from the data from \citet{yang_hyperbolicity_2009}. \REFone{Table~\ref{tab:dimension_comparison} collects the \NEW{model estimate} $d_M$, the autoencoder estimate $d_A$, the physical-mode count $d_{P}$, and the literature values $d_{\rm lit}$ for every $(L,\nu)$ studied.} The physical-mode counts $d_{P}$ shown in \fref{fig:d_vs_L_for_nu} and Table~\ref{tab:dimension_comparison} are obtained from the Lyapunov spectrum and are defined in the next section; we include them here for comparison.

\begin{table}
\centering
\begin{tabular}{c c c c c c}
$L$ & $\nu$ & $d_M$ & $d_A$ & $d_{P}$ & $d_{\rm lit}$ \\[2pt]
\hline
22  & 1.0  & 9   & $8\pm1$    & $8.0\pm0.3$    & 8  \\
44  & 1.0  & 18  & $19\pm1$   & $18.5\pm0.5$   & 18 \\
66  & 1.0  & 27  & $26\pm1$   & $28.5\pm0.5$   & 28 \\
96  & 1.0  & 39  & —          & —              & 41 \\
100 & 1.0  & 41  & $42\pm5$   & $41.0\pm1.0$   & —  \\
190 & 1.0  & 78  & —          & —              & 82 \\
200 & 1.0  & 82  & $90\pm5$   & $83.0\pm1.0$   & —  \\
\hline
22  & 0.1  & 29  & $30\pm5$   & $29.1\pm3.0$   & —  \\
44  & 0.1  & 57  & $60\pm5$   & $57.0\pm1.0$   & —  \\
66  & 0.1  & 86  & $85\pm2$   & $87.9\pm4.0$   & —  \\
100 & 0.1  & 130 & $130\pm10$ & $132.7\pm8.2$  & —  \\
200 & 0.1  & 260 & $260\pm10$ & $264.0\pm6.6$  & —  \\
\hline
22  & 0.01 & 90  & $90\pm10$  & $90.0\pm2.0$   & —  \\
44  & 0.01 & 181 & $180\pm10$ & $184.6\pm4.7$  & —  \\
66  & 0.01 & 271 & $280\pm10$ & $273.0\pm3.0$  & —  \\
100 & 0.01 & 411 & $410\pm10$ & $410.9\pm5.0$  & —  \\
\end{tabular}
\caption{\NEW{Model estimate} $d_M$ (rounded to the nearest mode), autoencoder estimate $d_A$, physical-mode count $d_{P}$ from the Lyapunov spectrum, and independent literature values $d_{\rm lit}$ (available only at $\nu=1$). The values at $L=96$ and $L=190$ are read from the figures of Ref.~\cite{yang_hyperbolicity_2009}
and have no corresponding autoencoder or Lyapunov estimate here.}
\label{tab:dimension_comparison}
\end{table}

\subsection{Lyapunov-based estimates}

As a second, independent estimate of the manifold dimension, we computed the full Lyapunov spectrum for every $(L,\nu)$ pair and, following \cite{yang_hyperbolicity_2009,takeuchi_hyperbolic_2011}, extracted the physical-mode count $d_{P}$ from the steep drop separating the near-zero exponents from the strongly damped modes. In every case the exponents stay close to zero up to the drop, whose location fixes $d_{P}$ and \NEW{agrees with the estimate} $d_M$. The estimates obtained this way are shown alongside $d_A$ and $d_M$ in \fref{fig:d_vs_L_for_nu}, where the three values agree across the full range of $L$ and $\nu$. The full set of spectra for all $(L,\nu)$, with the drop marked, is provided in Appendix~\ref{appB:further_examples}.

% \begin{figure}
%     \centering
%     \includegraphics[width=0.95\linewidth]{response_figs/fig_spectrum_L44_nu0.1_PM.pdf}
%     \caption{Lyapunov spectrum $\lambda_i$ for $L=44$, $\nu=0.1$. Orange dashed: prediction $d_M$; grey: physical-mode band read from the fall. \MELISA{Hay que subirla para que sea Fig. 6 y no Fig. 7, la dejo acá ahora.}}
%     \label{fig:lyap_L44_nu0.1}
% \end{figure}

\subsection{Comparison with the Fourier encoder}
\label{subsec:fourier_encoder}

To further substantiate our dimensional \NEW{estimate}, we trained a suite of hybrid models in which the nonlinear encoder was replaced by a fixed Fourier truncation, while the decoder remained a convolutional neural network. In this configuration, the decoder receives as input the first $n$ Fourier modes with wavenumber $k \le k_c$, where $k_c$ is a tunable cutoff. Each retained mode is represented by its real and imaginary components, so that the bottleneck dimension is $d = 2n$. For each choice of $k_c$ (and therefore each $d$), a separate network was trained using the same datasets and mean squared reconstruction loss as in the full autoencoder models. Again, implementation details can be found in \cite{repo_paper}.

As seen in \fref{fig:mse}, the Fourier encoder follows a similar pattern as the nonlinear autoencoder, where
the MSE drops sharply near $d=d_M$ and then plateaus. As a baseline for comparison, direct Fourier truncation (without a learned decoder) requires substantially more modes to attain similar error levels. \REFone{The same result holds across parameters: for every $(L,\nu)$ studied, retaining modes up to $k_d$ recovers the field while dropping below the cutoff does not (Appendix~\ref{appB:further_examples} shows more examples).}

A more qualitative view of the same behavior is given by the field reconstructions in \fref{fig:reconstruction_L44_nu0.1}, whose Fourier curves we deferred above. The top panel compares the original field with reduced representations obtained using the Fourier-truncated encoder with $k_c=k_d-1$ and $k_c=k_d$, as well as the full autoencoder. Retaining modes up to $k_d$ yields accurate reconstruction, whereas using only modes $k \le k_d - 1$ leads to noticeably larger errors and fails to reproduce the true field. Reconstructions based solely on Fourier filtering, even with the same number of retained modes, fail to recover the full signal. The lower panel shows the reconstruction errors $\Delta u(x)$ for this snapshot. \REFone{Consistent results were obtained for all configurations; the reconstructions for $\nu=1.0$ and $\nu=0.01$ are shown in Appendix~\ref{appB:further_examples}.} Interestingly, \cite{linot_data-driven_2022} reports similar findings using a Fourier encoder, although these results were not discussed in detail.

\section{Conclusions}

In summary, we propose a simple \NEW{model based on a} physical argument, to \NEW{estimate} the dimension of the inertial manifold of the Kuramoto–Sivashinsky equation. \NEW{In this model, the dimension} is determined by a spectral cutoff scale $k_d=\sqrt{5/(3\nu)}$ arising from energy balance considerations. This cutoff gives an explicit $d_M(L,\nu)$ for the number of dynamically active modes required to represent the long-time dynamics, predicting the extensive scaling with domain size. For $\nu=1.0$, the predicted values agree with previously calculated estimates. Using convolutional autoencoders and Lyapunov-based estimates, we independently determined the minimal latent dimension necessary to represent states on the attractor and find quantitative agreement with \NEW{this estimate}. Also, by training a decoder which takes Fourier modes as input, we provided evidence that \NEW{the Fourier modes up to $k_d$ suffice to parameterize the long-time dynamics}. \NEW{We stress, however, that our argument is heuristic: it does not constitute a rigorous result on the dimension of the inertial manifold, and establishing a rigorous connection between the energy-balance cutoff $k_d$ and the inertial manifold remains an open problem.}

Moving forward, several questions and issues remain open. Firstly, the argument proposed could be extended to the Complex Ginzburg Landau equation, where numerical estimates already exist, for a further test case. Secondly, the Navier Stokes (NS) case presents more challenges. There it was shown that the covariant Lyapunov vector analysis used for the KSE cannot be extended to the NS as the system is not hyperbolic \cite{inubushi_covariant_2012}. Also, autoencoder analysis of Kolmogorov flows \cite{cleary_estimating_2025,vinograd_dimensional_2026} and Rayleigh-Bénard flows \cite{Vinograd_Clark} show that classical mode-counting arguments overestimate the dimension of the manifolds calculated. Thirdly, understanding the difference between what the Kaplan-Yorke dimension and other estimates capture could provide interesting dynamical insights \cite{vinograd_dimensional_2026}. A related question is whether the dimension sufficient for reconstruction is also sufficient to close the dynamics in the latent space. Reconstruction-only autoencoders give no guarantee that the learned coordinates admit a simple dynamical model~\cite{lusch_deep_2018, champion_data-driven_2019}; for the KSE, the agreement of our static estimate with the dynamical Lyapunov count (\fref{fig:d_vs_L_for_nu}) suggests the two coincide here, and combining reconstruction with a dynamics-based loss is a natural way to test this directly. Lastly, and related to the last point, it is paramount to understand the Fourier structure of Lyapunov vectors, as this can shed light into the mechanisms involved \cite{yamada_Inertial_1988,Inubushi_Caulfield_2025}.

\begin{acknowledgments}
This work was partially supported by the Google PhD Fellowship Program. 
\end{acknowledgments}

%open questions:
%- relation to hyperbolicity
%   - CGL is hyperbollic, so it might work
%   - NS is not, but the argument is close to Landau's
%- relation to KY
%- Fourier analysis of lyapunov vectors

%Beyond the KSE, this mechanism suggests a general route to estimating inertial manifold dimensions in dissipative, spatially extended systems that admit finite-dimensional attractors and exhibit a clear separation between injection and dissipation scales, such as the complex Ginzburg–Landau equation. Although the Navier–Stokes equations do not admit such a simple closed-form spectral balance, our results highlight the central role of dynamically active spectral bands in determining effective dimensionality. This perspective is consistent with earlier analyses of Lyapunov spectra in turbulence \cite{yamada_Inertial_1988} and recent work relating dimensionality and synchronization in turbulent flows \cite{inubushi_dimensional_2025}.

\appendix
\section{Discrete vs. Continuous $k_d$}
\label{appA:discrete_modes}

In the main text, the cutoff wavenumber $k_d$ is derived using a continuum approximation for the Fourier modes. To quantify the error introduced by this approximation on the discrete grid used in our simulations, we evaluate the exact discrete balancing wavenumber $k_{d,\text{disc}} = N \Delta q$. Here, the discrete modes sit at $q_n = n \Delta q$ with a grid spacing of $\Delta q = 2\pi/L$ for $n = 1, 2, \dots, N$. The value $k_{d,\text{disc}}$ is found by locating the zero-crossing of the discrete sum:
\begin{equation}
  \sum_{n=1}^{N} \bigl(q_n^2 - \nu q_n^4\bigr) = 0 .
\end{equation}
Substituting $q_n = n \Delta q$ and factoring out $(\Delta q)^2$ yields:
\begin{equation}
  \sum_{n=1}^{N} n^2 - \nu (\Delta q)^2 \sum_{n=1}^{N} n^4 = 0 .
\end{equation}

Using the exact polynomial expressions for these power sums, we arrive at a reduced equation for the grid cutoff index $N$:

\begin{equation}
  \frac{1}{6} - \nu (\Delta q)^2 \frac{3N^2+3N-1}{30} = 0 .
\end{equation}

Multiplying by $30/\nu$ and grouping the remaining terms into $k_{d,\text{disc}} = N \Delta q$ gives:

\begin{equation}
  3 k_{d,\text{disc}}^2 + 3 \Delta q \, k_{d,\text{disc}} - (\Delta q)^2 = \frac{5}{\nu} .
\end{equation}

Dividing by 3 and identifying the continuous definition $k_d^2 = \frac{5}{3\nu}$, we obtain the quadratic relation:

\begin{equation}
  k_{d,\text{disc}}^2 + \Delta q \, k_{d,\text{disc}} - \frac{1}{3}(\Delta q)^2 = k_d^2 .
\end{equation}

Solving for the positive root via the quadratic formula reveals the exact relationship between the discrete and continuous balancing scales:
\begin{equation}
  k_{d,\text{disc}} = -\frac{\Delta q}{2} + \sqrt{k_d^2 + \frac{7}{12}(\Delta q)^2} \;\approx\; k_d - \frac{\Delta q}{2} + \mathcal{O}\left(\frac{(\Delta q)^2}{k_d}\right) .
\end{equation}

The continuum approximation $k_d$ overestimates the true discrete balancing point by roughly half a Fourier mode ($\Delta q / 2$). Correspondingly, the active dimension $d_M = 2\frac{L}{2\pi}k_d$ is overestimated by at most $1.0$ mode. This error is largest for our coarsest configuration, $L=22$, $\nu=1.0$, where the relative error is $\sim10.7\%$; it shrinks rapidly as the number of active modes grows. For $L=66$, $\nu=1.0$ it is $3\%$.

% As shown in Table~\ref{tab:kd_dm_relative_error}, the relative error is small even for our absolute coarsest configurations and rapidly vanishes as the number of active modes increases.

% As shown in Table~\ref{tab:kd_dm_relative_error}, the relative error is small even for our absolute coarsest configurations and rapidly vanishes as the number of active modes increases.

% \begin{table*}[ht]
% \centering
% \begin{tabular}{ccccccc}
% \hline
% $L$ & $\nu$ & $k_d$ (Theory) & $k_d$ (Discrete) & $d_M$ (Theory) & $d_M$ (Discrete) & Relative Error (\%) \\ \hline
% 22  & 1.00  & 1.29           & 1.17             & 9.04           & 8.17             & 10.67\%             \\
% 22  & 0.10  & 4.08           & 3.95             & 28.59          & 27.63            & 3.47\%              \\
% 22  & 0.01  & 12.91          & 12.77            & 90.41          & 89.42            & 1.10\%              \\
% 44  & 1.00  & 1.29           & 1.22             & 18.08          & 17.15            & 5.46\%              \\
% 66  & 1.00  & 1.29           & 1.25             & 27.12          & 26.16            & 3.66\%              \\ \hline
% \end{tabular}
% \caption{Comparison of theoretical (continuous) and exact discrete values.}
% \label{tab:kd_dm_relative_error}
% \end{table*}

\section{Further examples}
\label{appB:further_examples}

\begin{figure}[htbp]
    \centering
    \includegraphics[width=0.95\linewidth]{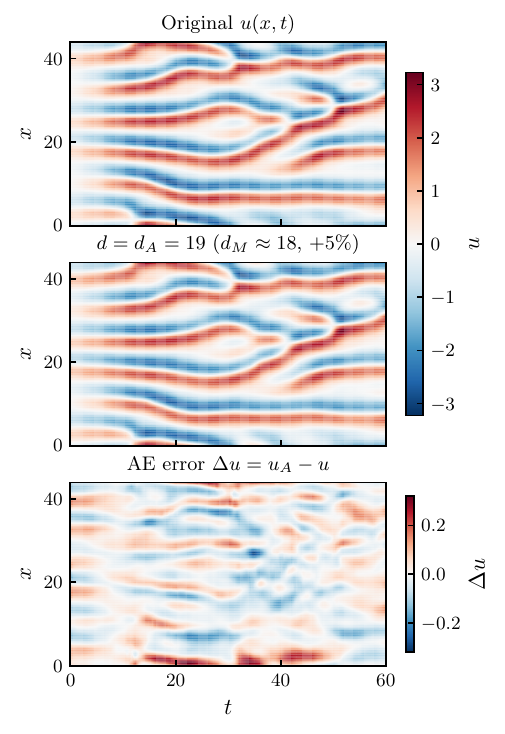}
    \caption{Space-time reconstruction for $L=44$, $\nu=1.0$. Panels as in
    \fref{fig:spacetime_L44_nu0.1}.}
    \label{fig:spacetime_L44_nu1}
\end{figure}

\begin{figure}[htbp]
    \centering
    \includegraphics[width=0.95\linewidth]{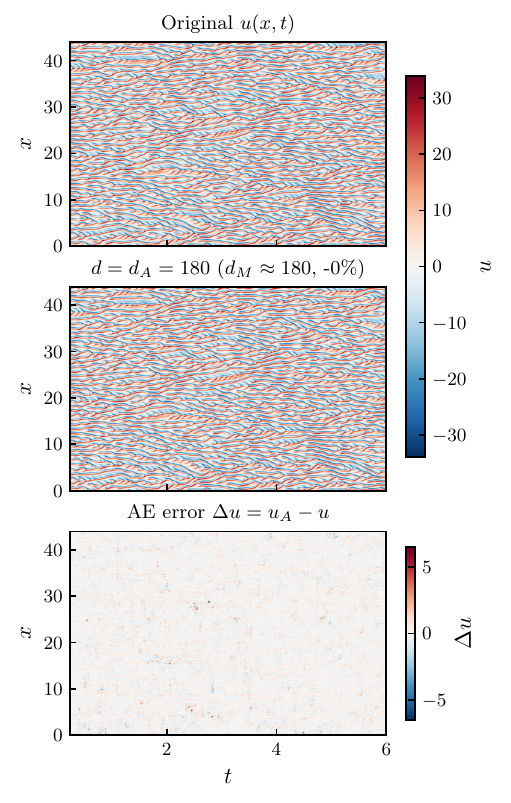}
    \caption{Space-time reconstruction for $L=44$, $\nu=0.01$. Panels as in
    \fref{fig:spacetime_L44_nu0.1}.}
    \label{fig:spacetime_L44_nu0.01}
\end{figure}

Figures~\ref{fig:spacetime_L44_nu1} and~\ref{fig:spacetime_L44_nu0.01} repeat the space-time comparison of \fref{fig:spacetime_L44_nu0.1} for $L=44$ at $\nu=1.0$ and $\nu=0.01$, spanning the least and most turbulent cases studied. In both the autoencoder reconstruction at $d\approx d_M$ reproduces the dynamics, and the error remains small and spatially uniform, with no accumulation on coherent structures.

\begin{figure*}
    \centering
    \includegraphics[width=0.95\linewidth]{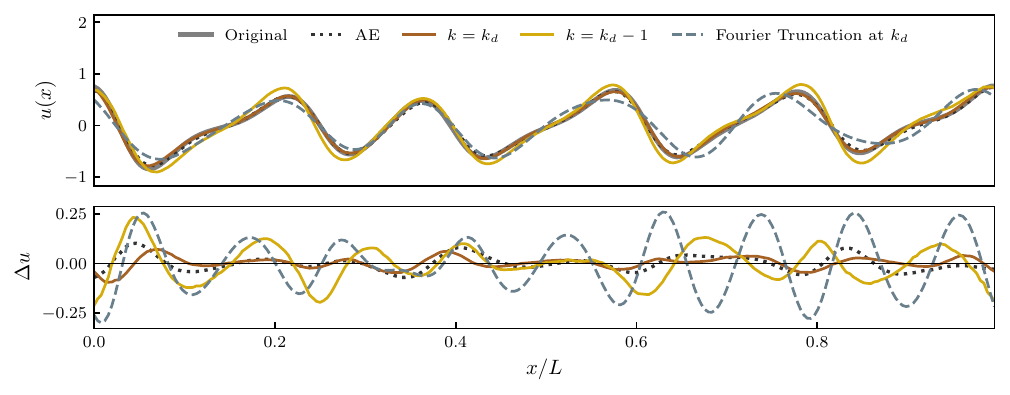}
    \caption{Single-snapshot reconstruction for $L=44$, $\nu=1.0$, shown over the full domain. Panels as in \fref{fig:reconstruction_L44_nu0.1}: top, original
    field and reduced representations; bottom, reconstruction errors $\Delta u(x)$.}
    \label{fig:reconstruction_L44_nu1}
\end{figure*}

\begin{figure*}
    \centering
    \includegraphics[width=0.95\linewidth]{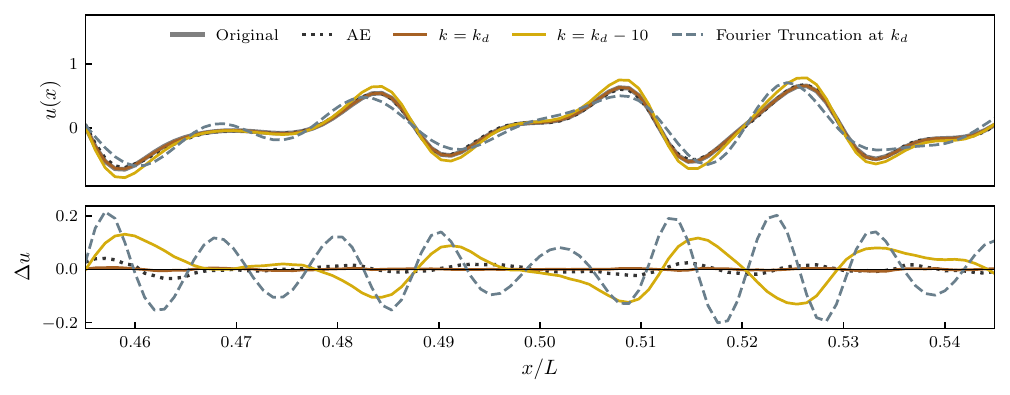}
    \caption{Single-snapshot reconstruction for $L=44$, $\nu=0.01$, shown over a subdomain to resolve the fine scales. Panels as in
    \fref{fig:reconstruction_L44_nu0.1}.}
    \label{fig:reconstruction_L44_nu0.01}
\end{figure*}

\fref{fig:reconstruction_L44_nu0.1} in the main text shows the single-snapshot reconstruction comparison for $L=44$, $\nu=0.1$. Figures~\ref{fig:reconstruction_L44_nu1} and~\ref{fig:reconstruction_L44_nu0.01} repeat it for $L=44$ at $\nu=1.0$ and $\nu=0.01$. The behavior is the same as at $\nu=0.1$: retaining modes up to $k_d$ (bottleneck $d=2n\approx d_M$) reconstructs the field accurately, dropping below the cutoff leaves visible errors, and direct Fourier truncation with the same number of modes fails to recover the signal.

\begin{figure}
    \centering
    \includegraphics[width=\linewidth]{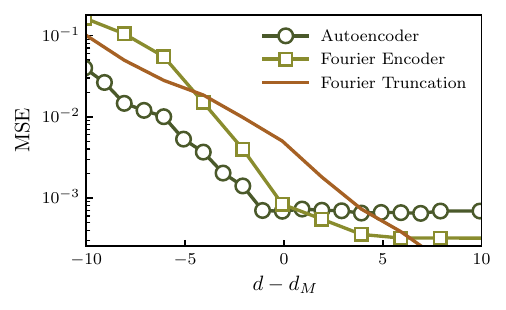}
    \caption{Mean–squared error (MSE) on the test data as a function of $d-d_M$ for $L=44$ and $\nu=1.0$, where $d_M$ is the \NEW{estimate from our model}. 
    }
    \label{fig:mse_L44_nu1.0}
\end{figure}

\begin{figure}
    \centering
    \includegraphics[width=\linewidth]{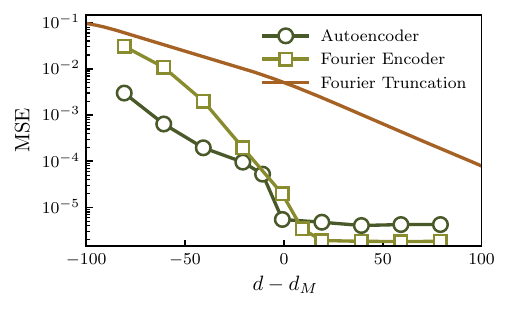}
    \caption{Mean–squared error (MSE) on the test data as a function of $d-d_M$ for $L=44$ and $\nu=0.01$, where $d_M$ is the \NEW{estimate from our model}. }
    \label{fig:mse_L44_nu0.01}
\end{figure}

The MSE-versus-dimension curves also extend across parameters. Figures~\ref{fig:mse_L44_nu1.0} and~\ref{fig:mse_L44_nu0.01} show the reconstruction MSE for $L=44$ at $\nu=1.0$ and $\nu=0.01$, complementing \fref{fig:mse} ($\nu=0.1$) in the main text. In both, the autoencoder and the Fourier encoder reach the plateau near $d=d_M$, while direct Fourier truncation requires substantially more modes to reach comparable error, showing that the spectral-cutoff threshold holds across the parameter range.

\begin{figure*}
    \centering
    \includegraphics[width=0.95\linewidth]{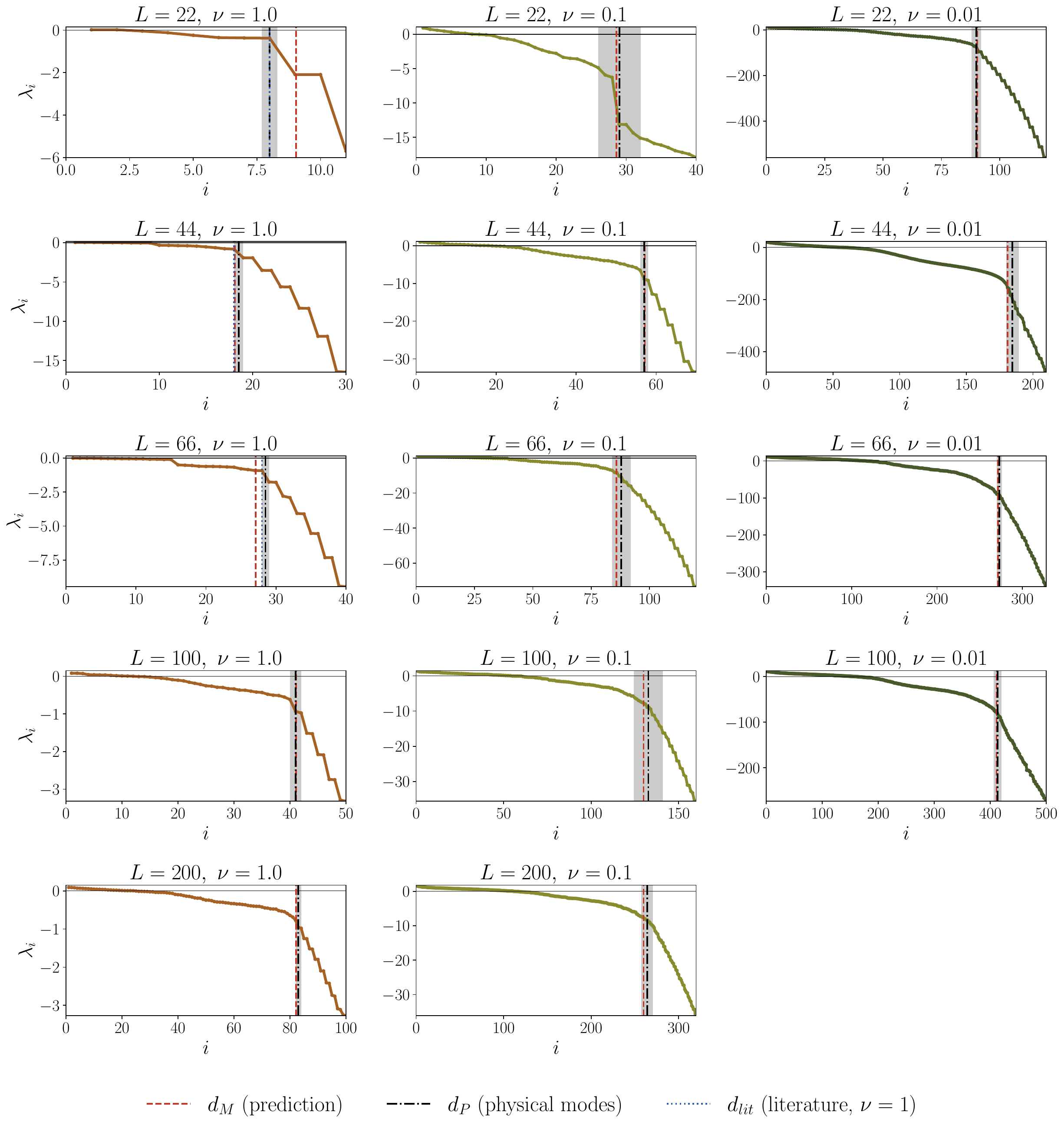}
    \caption{Lyapunov spectra $\lambda_i$ for all the $(L, \nu)$ studied. Orange dashed: \NEW{estimate} $d_M=(L/\pi)\sqrt{5/3\nu}$. Grey: physical-mode band read from the drop. Blue dotted ($\nu=1$ only): independent literature values $d_{\rm lit}$.}
    \label{fig:lyap_fall}
\end{figure*}

Figure~\ref{fig:lyap_fall} shows the ordered Lyapunov spectrum for every
$(L,\nu)$ pair considered in this work. As described in the main text, the
physical-mode count $d_{P}$ is extracted from the location of the steep drop
separating the near-zero exponents from the strongly damped transverse modes. At $\nu=1$, where independent literature values $d_{\rm lit}$ exist, the values we obtain coincides with $d_{\rm lit}$ and both agree with the \NEW{estimate} $d_M$ to within one mode. At $\nu=0.1$ and $\nu=0.01$, no literature values are available, but $d_M$ falls within the physical-mode band in every case.

\begin{figure}
    \centering
    \includegraphics[width=\linewidth]{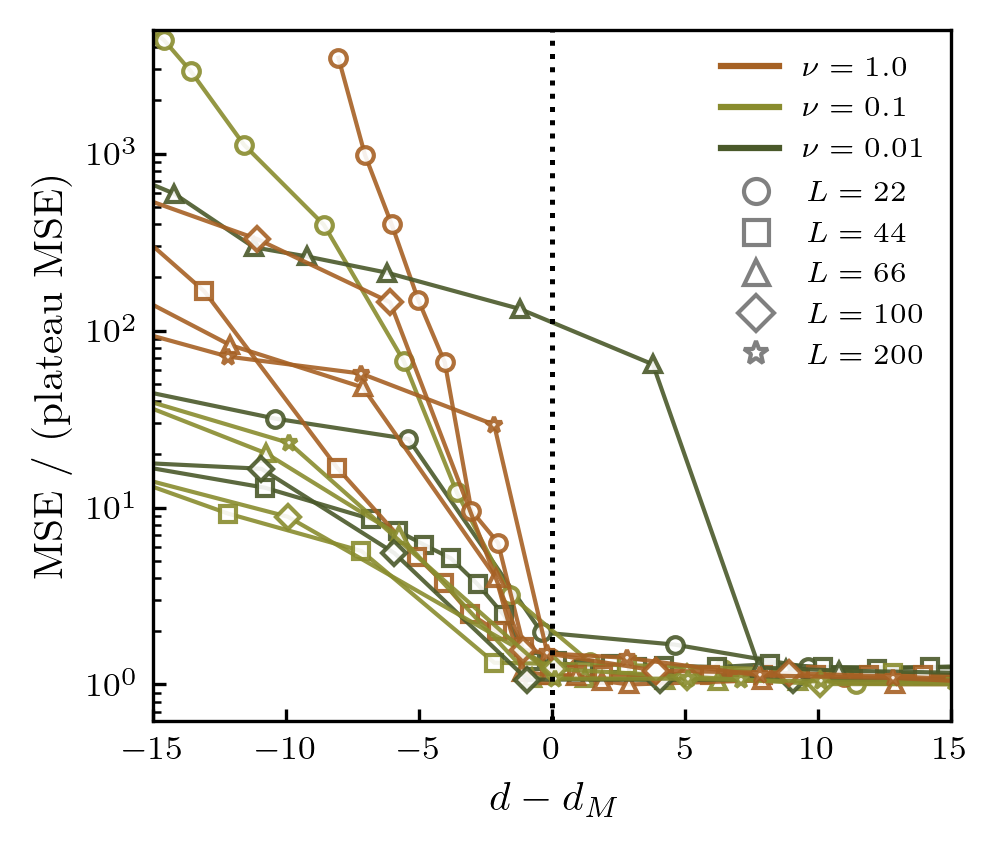}
    \caption{Autoencoder MSE normalized by its plateau value versus $d-d_M$ for all $(L,\nu)$ studied. Color denotes $\nu$, marker denotes $L$.}
    \label{fig:mse_collapse}
\end{figure}

Figure~\ref{fig:mse_collapse} shows the autoencoder reconstruction MSE, normalized by its plateau value, as a function of $d-d_M$ for all $(L,\nu)$ pairs studied. The MSE drops sharply as $d$ approaches $d_M$ and saturates for $d>d_M$ in every case, with the sharpness of the drop varying somewhat across parameters.

% Table~\ref{tab:dimension_comparison} collects the prediction $d_M$, the
% autoencoder estimate $d_A$, the physical-mode count $d_{P}$, and the literature values $d_{\rm lit}$ for every $(L,\nu)$ studied. The three estimates agree across the full range: at $\nu=1$, where literature values exist, $d_M$, $d_A$, and $d_{P}$ all fall on $d_{\rm lit}$ to within one mode, and at $\nu=0.1$ and $\nu=0.01$ the prediction and the two numerical estimates agree to a few percent.

% The relative discrepancies are largest where $d_M$ is smallest. This is expected: the prediction is a continuum mode-count, and Appendix~\ref{app:discrete} shows the discrete-grid correction is at most about one mode, which is a large fraction of a small $d_M$ ($11\%$ at $L=22$, $\nu=1$, where $d_M=9$) and negligible once $d_M$ is
% large. The physical-mode count carries its own uncertainty, reported as the spread of $d_{P}$ over Lyapunov integration times and the width of the fall.

\bibliography{bib}

\end{document}